\documentclass[review,12pt]{elsarticle}

\usepackage{lineno,hyperref}
\modulolinenumbers[5]

\journal{Carbon}

\begin{document}

\title{Outstanding strength, optical characteristics and thermal conductivity of graphene-like BC$_3$ and BC$_6$N semiconductors}

\author[buw,luh]{Bohayra Mortazavi\corref{cor1}}
\ead{bohayra.mortazavi@gmail.com}

\author[ru]{Masoud Shahrokhi}

\author[ikiu]{Mostafa Raeisi}

\author[luh]{Xiaoying Zhuang}

\author[ufrn]{Luiz Felipe C. Pereira\corref{cor1}}
\ead{pereira@fisica.ufrn.br}

\author[tus]{Timon Rabczuk}

\address[buw]{Institute of Structural Mechanics, Bauhaus-Universit\"at Weimar, Marienstr. 15, D-99423 Weimar, Germany}
\address[luh]{Cluster of Excellence PhoenixD (Photonics, Optics, and Engineering-Innovation Across Disciplines), 
Leibniz UniversitŠt Hannover, Hannover, Germany}
\address[ru]{Department of Physics, Faculty of Science, Razi University, Kermanshah, Iran}
\address[ikiu]{Mechanical Engineering Department, Imam Khomeini International University, Qazvin, Iran}
\address[ufrn]{Departamento de F\'{\i}sica, Universidade Federal do Rio Grande do Norte, Natal, 59078-970, Brazil}
\address[tus]{College of Civil Engineering, Department of Geotechnical Engineering, Tongji University, Shanghai, China}

\cortext[cor1]{Corresponding authors}

\date{\today}

\begin{abstract}
Carbon based two-dimensional (2D) materials with honeycomb lattices, like graphene, polyaniline carbon-nitride (C$_3$N) and boron-carbide (BC$_3$) exhibit exceptional physical properties. 
On this basis, we propose two novel graphene-like materials with BC$_6$N stoichiometry. 
We conducted first-principles calculations to explore the stability, mechanical response, electronic, optical and thermal transport characteristics of graphene-like BC$_3$ and BC$_6$N monolayers. 
The absence of imaginary frequencies in the phonon dispersions confirm dynamical stability of BC$_3$ and BC$_6$N monolayers. 
Our first principles results reveal that BC$_3$ and BC$_6$N present high elastic moduli of 256 and 305 N/m, and tensile strengths of 29.0 and 33.4 N/m, with room temperature lattice thermal conductivities of 410 and 1710 W/m.K, respectively. 
Notably, the thermal conductivity of BC$_6$N is one of the highest among all 2D materials.
According to electronic structure calculations, monolayers of BC$_3$ and BC$_6$N are indirect and direct bandgap semiconductors, respectively. 
The optical analysis illustrate that the first absorption peaks along the in-plane polarization for single-layer BC$_3$ and BC$_6$N occur in the visible range of the electromagnetic spectrum. 
Our results reveal outstandingly high mechanical properties and thermal conductivity along with attractive electronic and optical features of BC$_3$ and BC$_6$N nanosheets and present them as promising candidates to design novel nanodevices.
\end{abstract}


\maketitle

\section{Introduction}

Graphene \cite{Novoselov2004,Geim2007}, which is a two-dimensional (2D) carbon allotrope with a honeycomb atomic lattice, exhibits remarkable mechanical properties \cite{Lee2008} and ultrahigh thermal conductivity \cite{Balandin2008,Balandin2011,Xu2014} outperforming all known materials. Graphene also presents highly promising optical and electronic characteristics \cite{Berger2004,Liu2011a,Withers2010,Liu2019}. The exceptional properties of graphene not only propose this novel material for the design of a wide-variety of advanced devices, but also promoted the research for the design and synthesis of other 2D materials. Nonetheless, it is worthy to remind that for some critical technologies, pristine graphene does not fulfill the requirements. As a well-known example, for the application as nanotransistors in post-silicon electronics, presenting a direct and narrow band-gap semiconducting electronic character is essential, whereas graphene is a zero band-gap semimetal. To address this drawback, two main approaches have been extensively explored during the last decade. the first one includes the band-gap opening in graphene via defect engineering, mechanical straining, nanomesh creation or chemical functionalization \cite{Martins2007,Lherbier2012,Lherbier2008,Guinea2012,Pereira2009a,Guinea2010,Bai2010}. In these approaches, the opening of the band-gap in graphene requires additional processing steps after growth, which are complicated and expensive. Therefore a more appealing alternative that has been extensively explored during the last decade is to directly fabricate 2D semiconductors, such as C$_2$N \cite{Mahmood2015,Mortazavi2016a}, molybdenum disulfide \cite{Radisavljevic2011} and phosphorene \cite{Das2014,Li2014c} nanosheets.

Graphitic carbon nitride g-C$_3$N$_4$, layered materials have been widely synthesized for a long time by polymerization of cyanamide, dicyandiamide or melamine \cite{Thomas2008}. Graphitic carbon nitrides show porous atomic lattices and are made from covalent networks of carbon and nitrogen atoms. Unlike graphite, graphitic carbon nitrides are semiconductors. These layered materials have been proven as promising candidates for energy conversion and storage systems, catalysis, photocatalysis and oxygen reduction \cite{Thomas2008,Zheng2011,Lyth2011,Lyth2009,Makaremi2018}. Nevertheless, the first successful synthesis of large-area triazine-based graphitic carbon nitride nanosheets was reported in 2014 by Siller et al. \cite{Algara-Siller2014}, where it was produced on the basis of an ionothermal interfacial reaction. In 2015 another novel nanoporous carbon-nitride semiconducting nanosheet, so called nitrogenated holey graphene with a C$_2$N stoichiometry was fabricated by Mahmood et al. via a wet-chemical reaction \cite{Mahmood2015}. 
C$_2$N nanosheets and their C$_3$N$_4$ counterparts were theoretically predicted to be excellent candidates for photocatalysts \cite{Makaremi2018a,Zhang2018}. 
Shortly after, the same research group reported the first experimental realization of 2D polyaniline nanomembranes with a C$_3$N stoichiometry \cite{Mahmood2016}. Similarly to g-C$_3$N$_4$ and C$_2$N nanosheets, 2D polyaniline C$_3$N was found to be a semiconductor composed of carbon and nitrogen atoms only. Nonetheless, in contrast with g-C$_3$N$_4$ and C$_2$N, C$_3$N does not include a porous atomic lattice. The non-porous and densely packed atomic structure of C$_3$N nanosheets results in considerably higher mechanical properties and thermal conductivity in comparison with g-C$_3$N$_4$ and C$_2$N porous counterparts \cite{Mortazavi2017,Rajabpour2019,Dong2018,Gao2018,Shirazi2018,Sadeghzadeh2018,Shi2018,Hong2018}. C$_3$N graphene-like carbon nitride nanosheets have been proven to show desirable properties for numerous applications including nanotransistors \cite{Zhao2019,Zhang2019,Ren2019}, superconductivity \cite{Wang2019}, anode materials for Li-ion batteries \cite{Wang2019a}, and hydrogen storage \cite{Faye2019}. This shows that in recent years carbon nitride 2D semiconductors have attracted remarkable attention of theoretical and experimental research groups worldwide. 

Boron, like nitrogen, is also a neighbouring element of carbon with a very close atomic size and with the ability to form strong covalent bonds with it. This similarity raises questions concerning the stability and material properties of graphene-like boron carbide 2D nanostructures. 
In general, graphene-like materials made from B, C and N  show very attractive physical and chemical properties \cite{Thomas2019,Zhang2019a,Sreedhara2018,Zhang2018a,Zarei2018,Freitas2018}. 
Interestingly, more than a decade before the synthesis of C$_3$N nanomembranes \cite{Mahmood2016}, BC$_3$ layered sheets have been experimentally realized by Tanaka et al. via epitaxial growth on NbB$_2$ surfaces \cite{Tanaka2005}. 
Recently, graphene-like BC$_3$ nanosheets have been theoretically suggested as promising candidates for energy storage \cite{Qie2018}, nanoelectronics \cite{Tang2018,MehdiAghaei2018}, magnetic devices \cite{Chigo-Anota2016}, photocatalysts \cite{Zhang2018b} and catalysis \cite{Tang2018a}. 
Nevertheless, in spite of the much earlier experimental realization of BC$_3$ sheets, the available information concerning their intrinsic physical properties and application prospects are still limited when compared with its C$_3$N counterparts. 
In this work, we provide a comprehensive vision concerning the mechanical response, electronic, optical and thermal transport properties of graphene-like BC$_3$ monolayers via first-principles calculations. 
In addition to that, we predicted and explore the intrinsic properties of two graphene-like carbon-based nanomaterials  with BC$_6$N stoichiometry. 
These novel materials can be seen as transition structures between the C$_3$N and the BC$_3$ lattices. 
In fact, those novel direct band-gap semiconducting 2D materials not only yields high stiffness and attractive optical properties, but notably record some of the highest thermal conductivities among all predicted and fabricated 2D materials. 

\section{Computational methods}

Density functional theory (DFT) calculations in this work were performed employing the Vienna Ab-initio Simulation Package (VASP) \cite{Kresse1996,Kresse1996a,Kresse1999}. For the all simulations in this work, we used a plane-wave cutoff energy of 500 eV within the Perdew-Burke-Ernzerhof (PBE) generalized gradient approximation (GGA)  for the exchange correlation potential \cite{Perdew1996}. The convergence criteria for the electronic self consistence-loop was set to 10$^{-5}$ eV. To simulate nanosheets and not nanoribbons, periodic boundary conditions were applied along all three Cartesian directions, with a vacuum layer of 15 \AA~ to avoid image-image interactions along the monolayers thickness. The VESTA  package was used to illustrate atomic structures and charge densities \cite{Momma2011}. Energy minimized BC$_3$ and BC$_6$N monolayers were obtained by altering the size of the hexagonal unit-cells and subsequently performing geometry optimizations of the atomic positions, employing the conjugate gradient method. The convergence criteria for the HellmannÐFeynman forces on each atom was taken to be 0.01 eV/\AA~ employing a 15x15x1 Monkhorst-Pack \cite{Monkhorst1976} k-point mesh. After obtaining energy minimized lattices, uniaxial tensile simulations were carried out to explore the mechanical properties. The electronic properties were evaluated using a denser k-point grid of 21x21x1. Since PBE/GGA underestimates the band-gap values, we employed the screened hybrid functional HSE06 \cite{Krukau2006} to provide more accurate estimations. Optical properties were analyzed on the basis of the random phase approximation (RPA) constructed over the PBE results. 
Thermal stability of the considered nanosheets was examined via ab-initio molecular dynamics simulations (AIMD) for 2x2x1 super-cells with the Langevin thermostat, a time step of 1 fs and a 2x2x1 Monkhorst-Pack k-point mesh size \cite{Monkhorst1976}.

The phononic thermal conductivities of single-layer BC$_3$ and BC$_6$N were predicted with the ShengBTE package \cite{Li2014}, which conducts a fully iterative solutions of the Boltzmann transport equation. Further details of the thermal conductivity calculations can be found in a previous study concerning the C$_3$N monolayer \cite{Peng2018}. Second order (harmonic) and third-order (anharmonic) interatomic force constants were calculated using density functional perturbation theory (DFPT)  as implemented in the VASP package, also on the basis of PBE/GGA and for 4x4x1 super-cells with 3x3x1 k-point grids. 
Phonon frequencies, group velocities and harmonic interatomic force constants were obtained with PHONOPY \cite{Togo2015}, from inputs provided by the DFPT results. In accordance with the prebious study \cite{Peng2018}, for the third-order anharmonic force constants, interactions up to the eleventh nearest-neighbours were considered. 
Born effective charges and dielectric constants were also considered in the dynamical matrix to obtain the thermal conductivity with a 51x51x1 q-point mesh. Nonetheless, we found that aforementioned terms can be accurately neglected for the studied nanosheets as their contributions in the estimated thermal conductivities are below 1\%.

\section{Results and discussion}

\begin{figure}[htbp]
\begin{center}
\includegraphics[width=\linewidth]{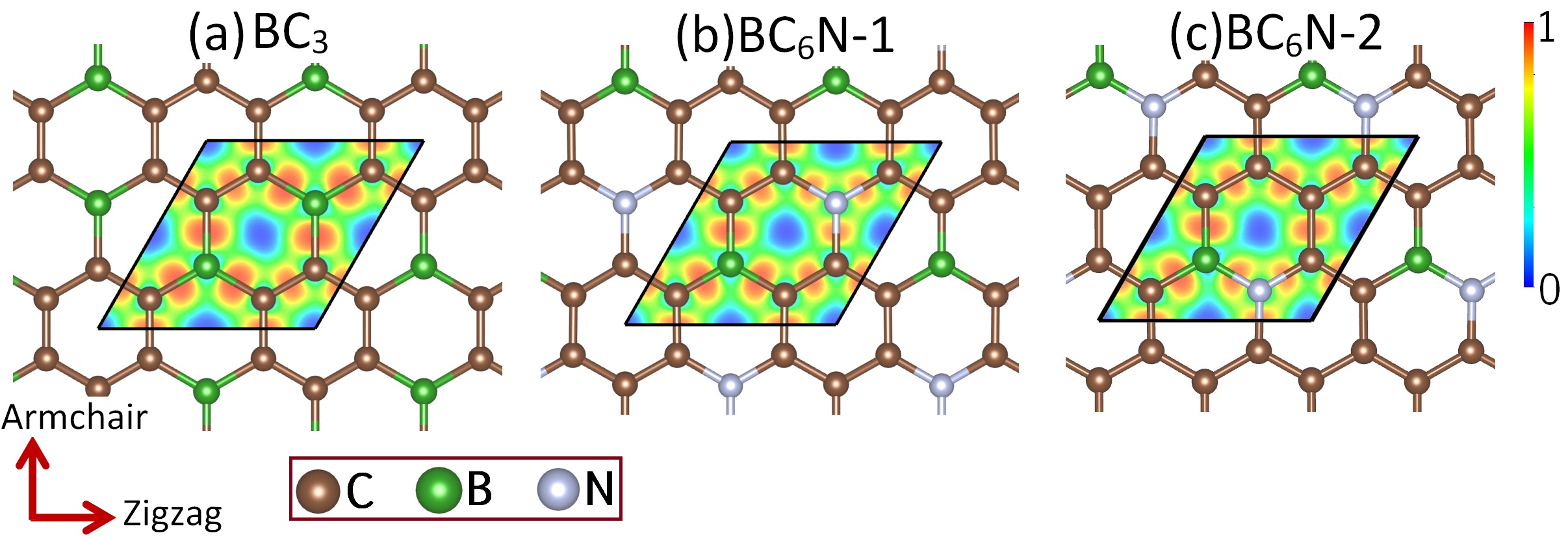}
\caption{Energy minimized atomic structure of BC$_3$, BC$_6$N-1 and BC$_6$N-2 monolayers. Contours illustrate the electron localization function within the unit-cell \cite{Silvi1994}.}
\label{fig01}
\end{center}
\end{figure}

We begin by pointing out that BC$_3$ and C$_3$N nanosheets have been experimentally fabricated, thus the considered atomic lattices are the most stable structures. With respect to the BC$_6$N structure, considering a hexagonal unit-cell with 8 atoms, only two different structures can be formed. 
These two structures include lattices without a B-N bond and with a single B-N bond, which we identify as BC$_6$N-1 and BC$_6$N-2, respectively.
Energy minimized BC$_3$, BC$_6$N-1 and BC$_6$N-2 monolayers with graphene-like hexagonal atomic lattices are illustrated in Fig. \ref{fig01}. 
The lattice constants of BC$_3$, BC$_6$N-1, BC$_6$N-2 and C$_3$N monolayers were estimated to be 5.174, 4.979, 4.973 and 4.860 \AA, respectively. In single-layer BC$_3$, the C-C and C-B bond lengths were measured to be 1.565 and 1.422 \AA, respectively, which are longer than C-C and C-N bonds in the C$_3$N counterpart ($\approx$ 1.403 \AA). 
For the BC$_6$N-1 monolayer, C-C, C-B and C-N bonds were found to be 1.413, 1.471 and 1.453 \AA, respectively.
In the case of the BC$_6$N-2 monolayer, B-N, C-B and C-N bonds were measured to be 1.453, 1.410 and 1.486 \AA, respectively.
The lattice energy per atom for BC$_6$N-1 and BC$_6$N-2 monolayers, were predicted to be -8.71 and -8.85 eV, respectively, which indicate that BC$_6$N-2 should be slightly more stable than BC$_6$N-1.
To facilitate future studies, the unit-cells of energy minimized monolayers are provided in supplementary information. 
In analogy to graphene and in order to investigate the anisotropy in the mechanical and thermal conduction responses, armchair and zigzag directions can be defined for the studied nanosheets, as shown in Fig. \ref{fig01}. 
To provide useful insights concerning the atomic bonding nature in the studied nanosheets, the electron localization function (ELF)  \cite{Silvi1994} within the unit-cells is also illustrated in the figure. ELF is a spatial function and takes a value between 0 and 1. As expected, electron localization occurs around the center of all bonds in these nanomembranes, revealing the dominance of covalent bonding between pairs of atoms. 
Interestingly, for the all considered monolayers the electron localization around the center of connecting bonds is broadened for C-B bonds in comparison with C-N and C-C bonds. In this case, for C-N bonds the electron localization shows the most concentrated pattern, which might be an indication of the higher stiffness of these covalent bonds. 

\begin{figure}[htbp]
\begin{center}
\includegraphics[width=0.9\linewidth]{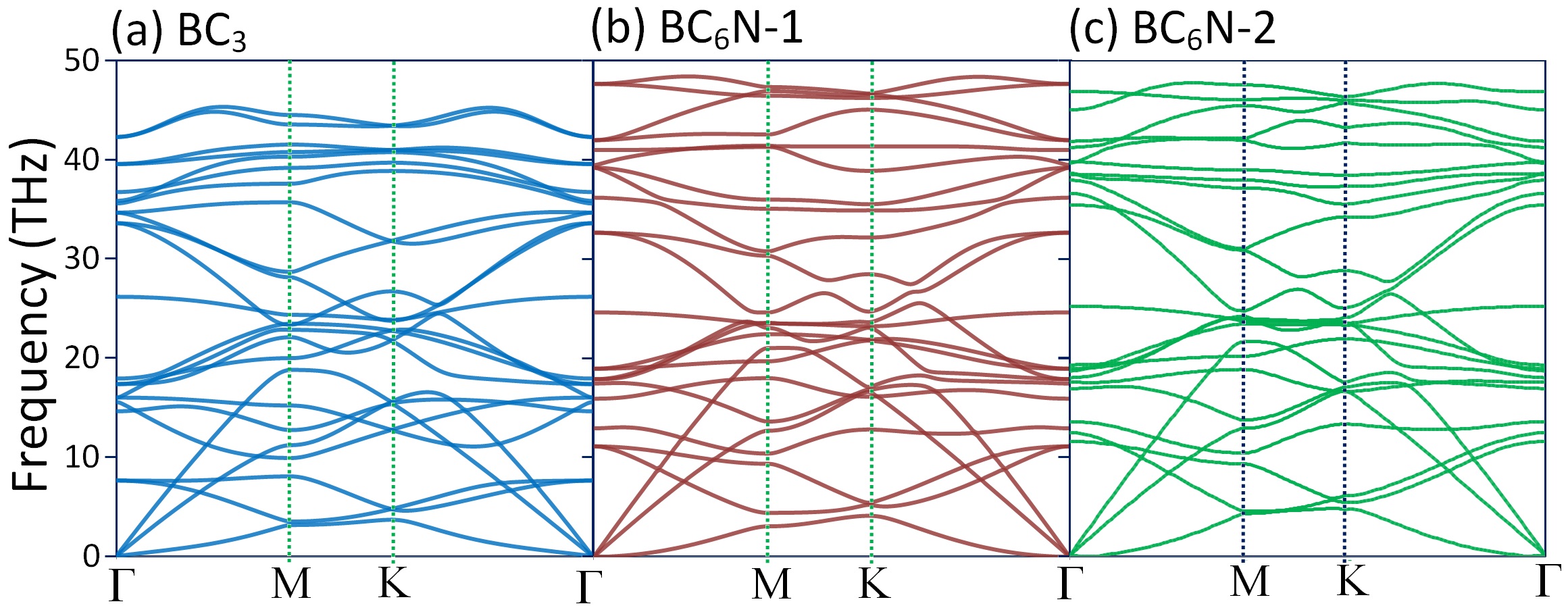}
\caption{Phonon frequencies for BC$_3$, BC$_6$N-1 and BC$_6$N-2 along high symmetry directions of the first Brillouin zone. Notice the presence of two acoustic modes with linear dispersion while the remaining one presents a quadratic dispersion characteristic of monolayer 2D materials.}
\label{fig02}
\end{center}
\end{figure}

Phonon frequencies for BC$_3$, BC$_6$N-1 and BC$_6$N-2 along high symmetry directions of the first Brillouin zone are shown in Fig. \ref{fig02}. In all cases, two of the three acoustic modes present linear dispersion while the remaining one presents a quadratic dispersion, which is characteristic of monolayer 2D materials. 
Around the M point the transverse and longitudinal acoustic modes reach frequencies around 10 and 20 THz, respectively, whereas the flexural mode remains below 5 THz.
A visual inspection of the slope of the linear acoustic modes in the phonon dispersions of Fig. \ref{fig02} indicates a lower speed of sound for BC$_3$ in comparison to the BC$_6$N structures. 
This will also influence the thermal conductivities as we shall see later.
The absence of imaginary eigenvalues for the dynamical matrix, which would appear in the figure as negative frequencies, is a good indication of the structural stability for both structures. 
Since both BC$_3$ and C$_3$N have already been synthesized, the predicted structural stability of both BC$_6$N should serve as encouragement to those with an interest in producing those novel 2D materials.

\begin{figure}[htbp]
\begin{center}
\includegraphics[width=0.9\linewidth]{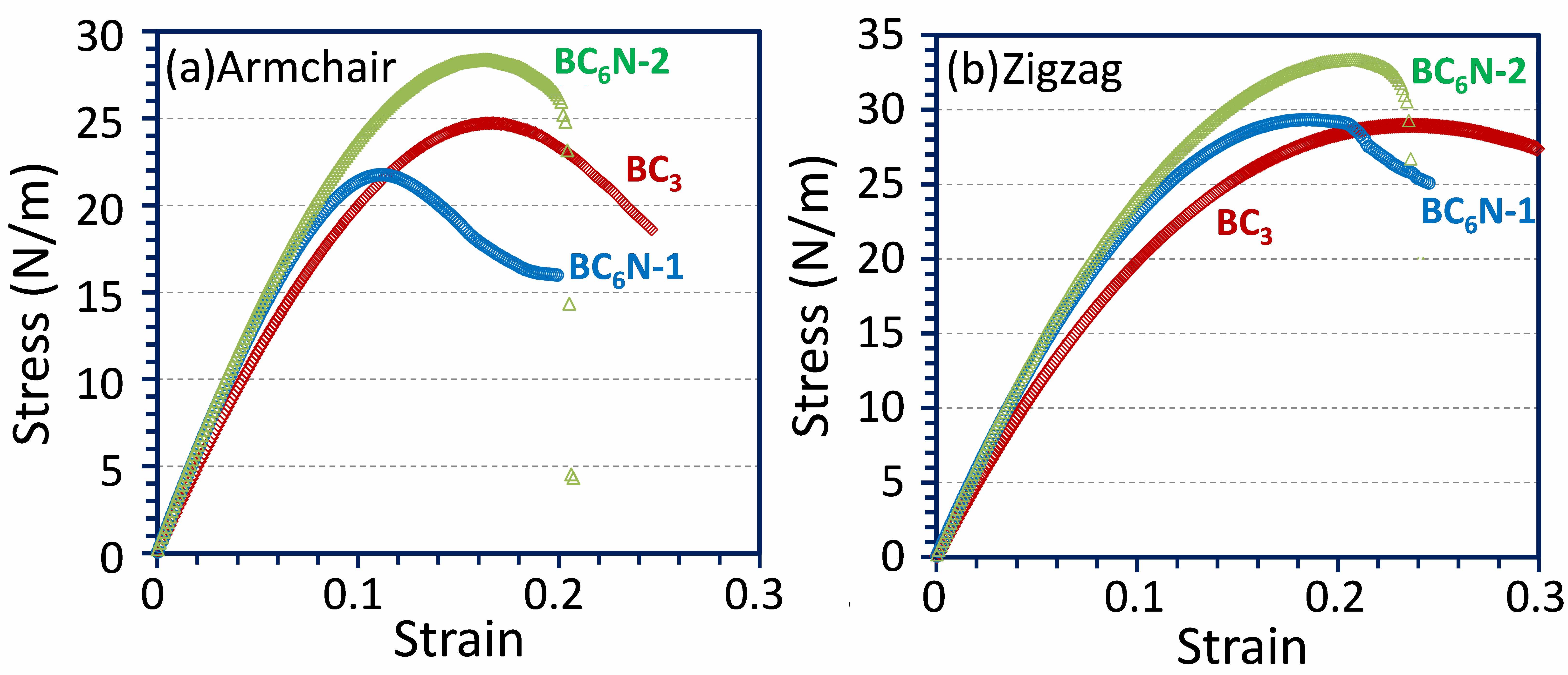}
\caption{Uniaxial stress-strain response of BC$_3$, BC$_6$N-1 and BC$_6$N-2 monolayers elongated along (a) armchair and (b) zigzag directions. All three materials present isotropic elastic response along the in-plane directions.}
\label{fig03}
\end{center}
\end{figure}

Mechanical properties of BC$_3$ and BC$_6$N nanosheets have been investigated by conducting uniaxial tensile simulations. In order to check for a possible anisotropy in the mechanical response, uniaxial tensile simulations were conducted along armchair and zigzag directions. During uniaxial tensile loading, the periodic dimension along the loading direction was increased step-by-step with a fixed strain of 0.0005. In order to satisfy the uniaxial stress-conditions, the dimension perpendicular to the loading direction was adjusted to reach a negligible stress ($<$0.03 N/m). The atomic positions were rescaled according to the changes in the simulation box size, and subsequently energy minimization was conducted within the conjugate gradient method in order to allow the rearrangement of atomic positions.  In Fig. \ref{fig03} the DFT prediction for the uniaxial stress-strain response of BC$_3$, BC$_6$N-1 and  BC$_6$N-2 nanosheets along armchair and zigzag directions are compared. As expected, the curves exhibit an initial linear behavior, corresponding to the elastic region. For each monolayer, these linear regions along armchair and zigzag directions were found to coincide, revealing isotropic elastic response for BC$_3$, BC$_6$N-1 and  BC$_6$N-2 monolayers. Moreover, we find that BC$_6$N-1 and  BC$_6$N-2 have the same elastic moduli.

The elastic moduli of BC$_3$ and BC$_6$N atomic layers were predicted to be considerably high, 256 and 305 N/m, respectively, which are lower than that of the C$_3$N counterpart $\approx$ 340 N/m \cite{Mortazavi2017}. These results highlight the stiffening role of C-N bonds when compared with C-B ones. Within the elastic range, the strain along the traverse direction of loading ($s_t$) with respect to the loading strain ($s_l$) is constant and can be used to evaluate  Poisson's ratio $-s_t/s_l$. The Poisson's ratios of BC$_3$ and BC$_6$N monolayers were estimated to be 0.180 and 0.175, respectively. According to our DFT results, BC$_3$ and BC$_6$N nanosheets exhibit distinctly higher tensile strength and strain at tensile strength point (which is a representative of stretchability) along the zigzag direction when compared with the armchair direction. Therefore, the tensile response of these graphene-like lattices are dependent on the loading directions and thus anisotropic. 
However, BC$_3$ shows  a lower degree of anisotropy due to its more uniform atomic configuration, with tensile strengths of 29.0 and 24.7 N/m at corresponding strain values of 0.235 and 0.165 in the case of uniaxial loading along zigzag and armchair directions, respectively. 
Meanwhile, in the case of BC$_6$N-1 monolayer, we predict tensile strengths of 21.8 and 29.3 N/m at corresponding strain values of 0.11 and 0.19 for the uniaxial loading along armchair and zigzag directions, respectively. 
Finally, among the considered structures, BC$_6$N-2 exhibits the highest tensile strengths of 28.4 and 33.4 N/m at corresponding strain values of 0.17 and 0.21 the armchair and zigzag directions, respectively.
These results  confirm the outstandingly high tensile strength of BC$_3$ and BC$_6$N monolayers.

Let us point out that the remarkable tensile strength and elastic moduli of BC$_3$ and BC$_6$N nanosheets are not enough to ensure their thermal stability. 
Therefore, in order to probe their thermal stability we conducted AIMD simulations at 500 K and 1000 K for a total simulation time of 20 ps. 
The results are presented in the Supporting Information, and according to Fig. S1, BC$_3$ and both BC$_6$N atomic lattices were kept intact even at the high temperature of 1000 K, which is a strong confirmation of their thermal stability.

\begin{figure}[htbp]
\begin{center}
\includegraphics[width=\linewidth]{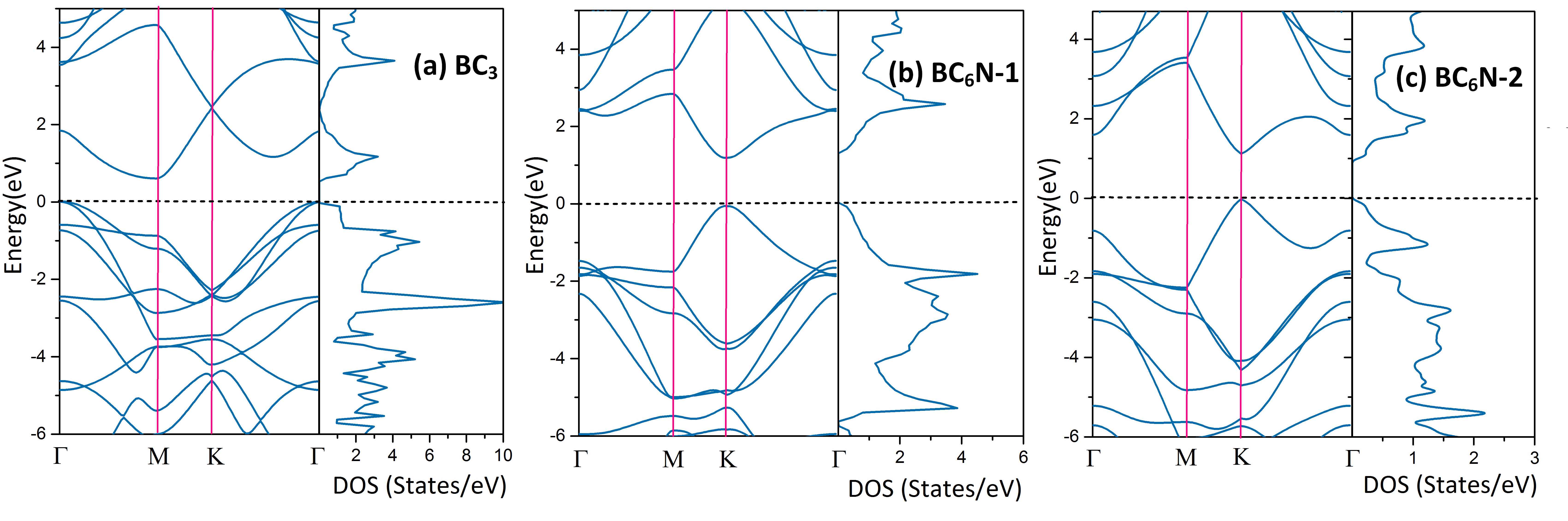}
\caption{Band structure and total electronic density of states (DOS) of BC$_3$, BC$_3$, BC$_6$N-1 and  BC$_6$N-2 monolayers predicted by the PBE/GGA functional. The Fermi energy is aligned to zero}
\label{fig04}
\end{center}
\end{figure}

Now we investigate the electronic and optical characteristics of BC$_3$, BC$_6$N-1 and  BC$_6$N-2 monolayers. We begin by obtaining the electronic band structure along high symmetry directions of the first Brillouin zone as shown in Fig. \ref{fig04}. 
Our results, based on PBE/GGA, show that the valence band maximum (VBM) in BC$_3$ monolayer occurs at the $\Gamma$ point while the conduction band minimum (CBM) occurs at M-point, exhibiting therefore an indirect bandgap semiconducting character. 
These results also show that both the VBM and CBM of BC$_6$N-1 and  BC$_6$N-2  occur at the K-point, resulting in a direct bandgap. 
In comparison, C$_3$N monolayer is an indirect bandgap semiconductor, since its VBM and CBM are located at M- and $\Gamma$-points, respectively \cite{Makaremi2017}.  
According to the PBE results, the bandgaps of BC$_3$, BC$_6$N-1 and  BC$_6$N-2 monolayers are 0.62,1.26 and 1.14 eV, respectively. 
It is a known issue that DFT within the PBE/GGA level of theory underestimates the bandgap of semiconductors \cite{Sham1983}, therefore we employ the HSE06 method to provide more accurate predictions. 
The corresponding bandgaps (shown in Fig. S2) for BC$_3$, BC$_6$N-1 and  BC$_6$N-2 monolayers within HSE06 functional are 1.82, 2.10 and 1.77 eV, which are indeed larger than those predicted by PBE/GGA. 
In any case, the predicted semiconducting character of BC$_3$ and BC$_6$N presents an advantage relative to graphene's zero bandgap semimetallic behavior, at least when it is necessary to switch the conductivity between on and off states. 
It is worth noting that total electronic density of states (DOS) were also acquired from spin-polarized calculations. The DOS for spin-up and spin-down channels were found to be completely symmetrical and free of spin-splitting, confirming the non-magnetic semiconducting electronic character of BC$_3$, BC$_6$N-1 and BC$_6$N-2 nanosheets.

\begin{figure}[htbp]
\begin{center}
\includegraphics[width=0.9\linewidth]{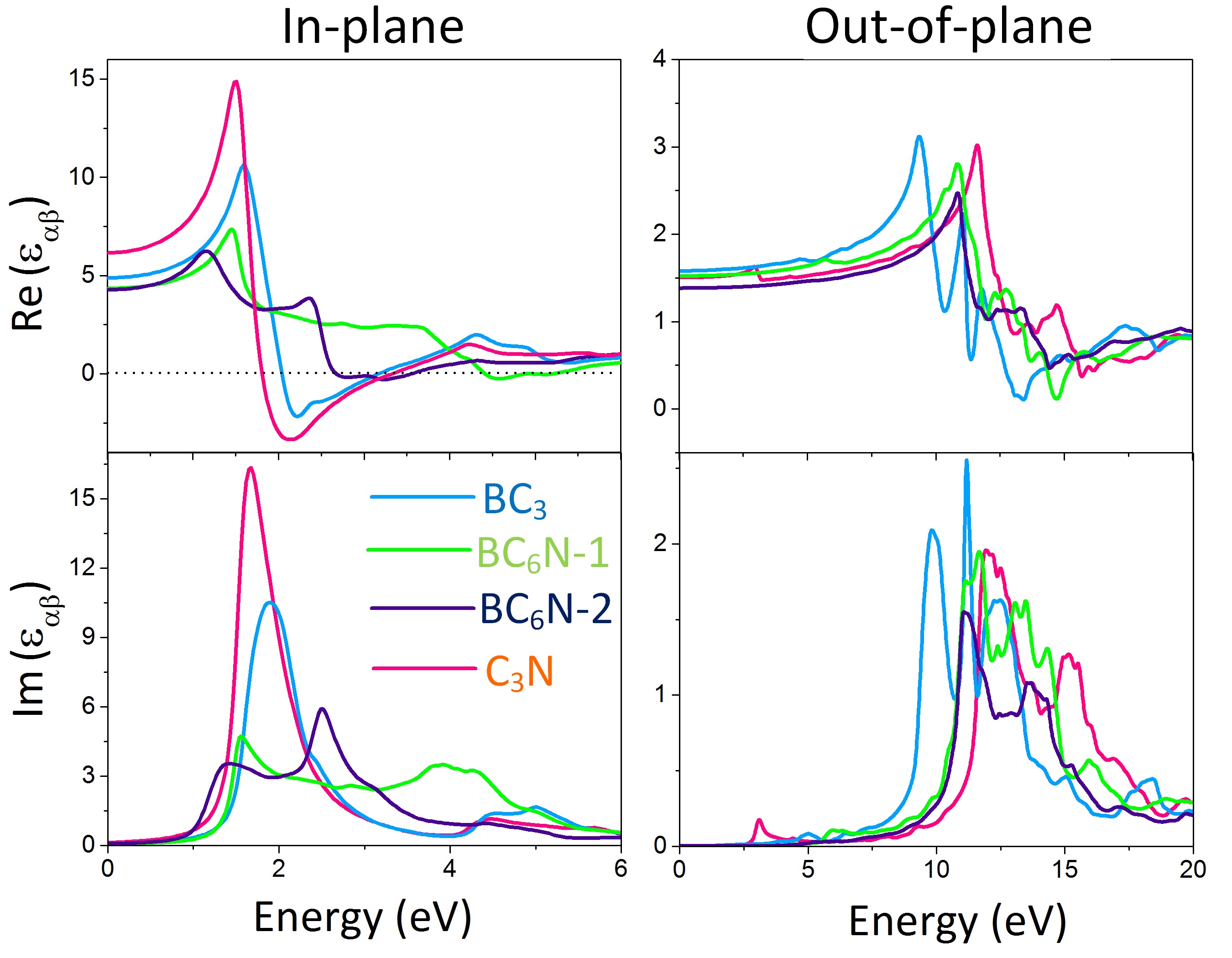}
\caption{Imaginary and real parts of the dielectric function of single-layer BC$_3$, BC$_6$N and C$_3$N for in-plane and out-of-plane polarizations, calculated using the RPA+PBE approach}
\label{fig05}
\end{center}
\end{figure}

Once the electronic structure calculations confirmed the semiconducting character of BC$_3$ and BC$_6$N monolayers, we probed their optical properties for possible applications in optoelectronics. 
In this case, calculations were also conducted for the C$_3$N monolayer in order to provide a more comprehensive vision. 
Imaginary and real parts of the dielectric function, $\mathrm{Im}[\epsilon_{\alpha \beta}]$ and $\mathrm{Re}[\epsilon_{\alpha \beta}]$, for BC$_3$, BC$_6$N-1, BC$_6$N-2 and C$_3$N sheets are presented in Fig. \ref{fig05} as functions of photon energy. 
We consider parallel (in-plane) and perpendicular (out-of-plane) polarization directions within RPA+PBE. 
In the case of in-plane polarization, the absorption edge of $\mathrm{Im}[\epsilon_{\alpha \beta}]$, which corresponds to the optical gap, occurs at 1.45, 1.39, 1.11 and 1.27 eV for BC$_3$, BC$_6$N-1, BC$_6$N-2 and C$_3$N respectively. 
The corresponding values along the out-of-plane direction are 4.57, 5.5, 5.22 and 2.88 eV, respectively. 
The first main peaks of $\mathrm{Im}[\epsilon_{\alpha \beta}]$ along the in-plane polarization for the all considered monolayers happen in the visible range, and are related to $\pi \rightarrow \pi^*$ transitions. 
In the case of BC$_3$ the prominent $\mathrm{Im}[\epsilon_{\alpha \beta}]$ peak along in-plane polarization has a blue shift when compared with BC$_6$N and C$_3$N. 
The main peaks along the out-of-plane direction for these nanosheets are broad and occur in energy range between 9.32 and 13.00 eV, being  related to $\pi \rightarrow \sigma^*$ and $\sigma \rightarrow \pi^*$  transitions.

\begin{figure}[htbp]
\begin{center}
\includegraphics[width=0.9\linewidth]{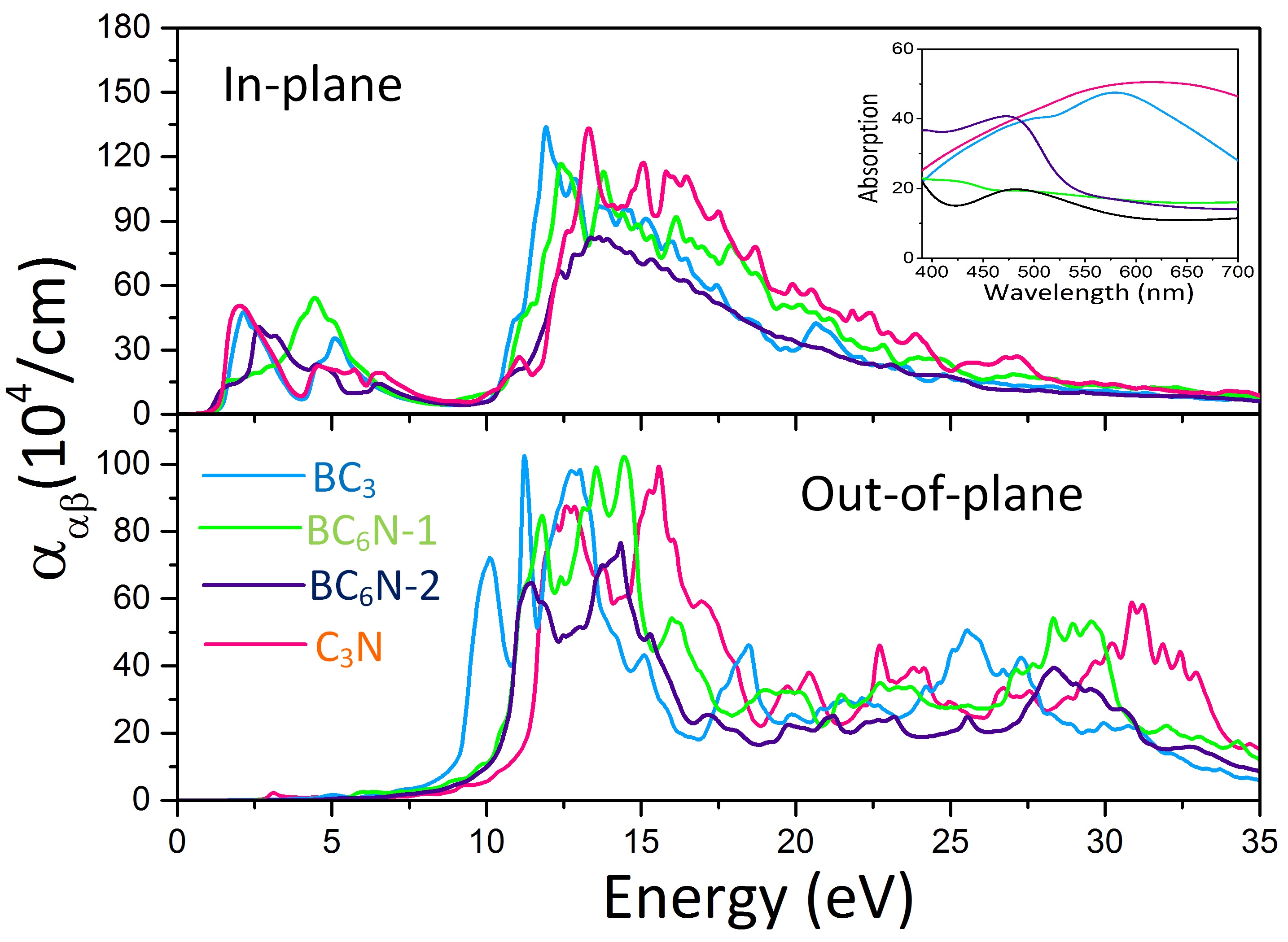}
\caption{Optical absorption spectra $\alpha_{\alpha \beta}(\omega)$ of single-layer BC$_3$, BC$_6$N and C$_3$N  as a function of photon energy for in-plane and out-of-plane polarizations within RPA+PBE approach. The inset shows a comparison of absorption spectra for the in-plane polarization in the visible range of light as a function of wavelength including graphene (black line)}
\label{fig06}
\end{center}
\end{figure}

The static dielectric constants (real part of the dielectric constant at zero energy) for BC$_3$, BC$_6$N-1, BC$_6$N-2 and C$_3$N were calculated to be 4.86, 4.35, 4.22 and 6.16 for in-plane polarization and 1.59, 1.54, 1.38  and 1.49 for out-of-plane polarization.
 For BC$_3$, BC$_6$N-1, BC$_6$N-2 and C$_3$N the main peak of $\mathrm{Re}[\epsilon_{\alpha \beta}]$ along the in-plane direction occur at 1.60, 1.44, 1.14 and 1.49 eV while the corresponding values for out-of-plane polarization were found to be 9.32, 10.88, 10.90 and 11.62 eV, respectively. 
 In what follows, we discuss the absorption coefficient, $\alpha_{\alpha \beta}(\omega)$, which is given by \cite{wooten1972optical}
\begin{equation}
\alpha_{\alpha \beta}(\omega) = \frac{\omega \, \mathrm{Im} [\epsilon_{\alpha \beta} (\omega)]}{c \, n_{\alpha \beta} (\omega)},
\end{equation}
where $c$ is the speed of light and $n_{\alpha \beta} (\omega)$ is the refraction index. 
The absorption coefficients for  BC$_3$, BC$_6$N-1, BC$_6$N-2 and C$_3$N are plotted in Fig. \ref{fig06}. In this case we also compared the acquired results with that of pristine graphene in the visible range of light (from 390 to 700 nm) as a function of wavelength from our previous studies \cite{Shahrokhi2016,Shahrokhi2016a}. 
The inset of Fig. \ref{fig06} shows the absorption coefficients of single-layer BC$_3$, BC$_6$N-1, BC$_6$N-2 and C$_3$N and graphene (black line) in the visible range. 
These results indicate that the first absorption peaks for BC$_3$, BC$_6$N-1, BC$_6$N-2 and C$_3$N monolayers are 2.13, 1.66, 1.55 and 2.01 eV along the in-plane polarization, respectively, which are indeed in the visible range. 
The main peaks of $\alpha_{\alpha \beta}(\omega)$ for BC$_3$, BC$_6$N-1, BC$_6$N-2 and C$_3$N along in-plane polarization were found to be broad and located at an energy range between 10.78 and 18.00 eV. 
The first absorption peaks along the out-of-plane polarization occur at 5.04, 5.89, 5.30 and 3.10 eV, respectively, which are in the ultraviolet (UV)  range of light, while the main absorption peaks in this direction locate at energy levels between 9.19 and 18.15 eV. 
As shown in the inset of Fig. \ref{fig06} the absorption coefficients for BC$_3$, BC$_6$N-1, BC$_6$N-2 and C$_3$N monolayers in the visible range of light are higher than that of graphene. These results indicate that these monolayers can enhance visible-light absorption in comparison with graphene, which can be potentially attractive for photovoltaic applications. 

\begin{figure}[htbp]
\begin{center}
\includegraphics[width=0.9\linewidth]{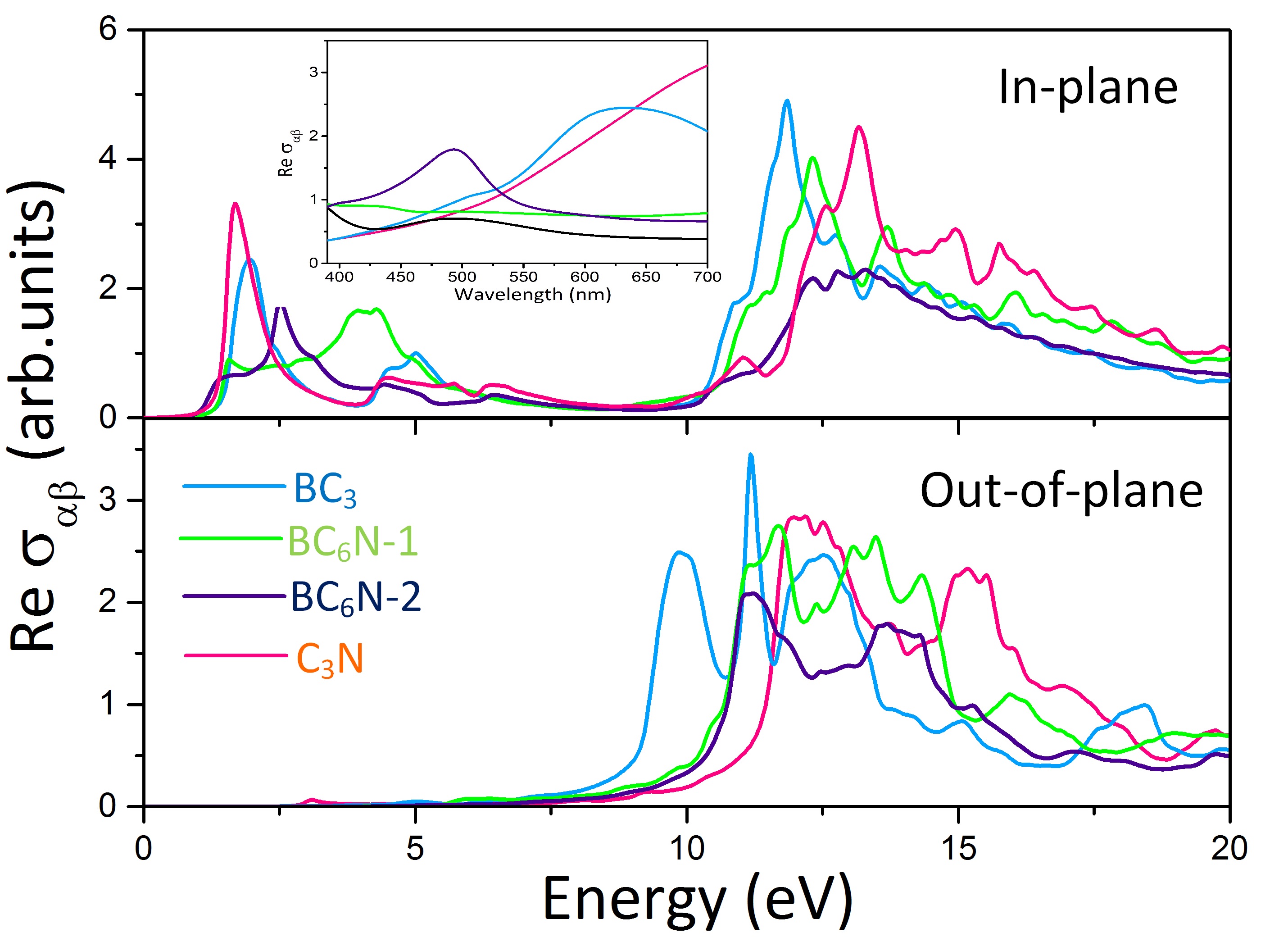}
\caption{Optical conductivity, $\mathrm{Re}[ \sigma_{\alpha \beta}(\omega)]$, of single-layer BC$_3$, BC$_6$N and C$_3$N  as a function of photon energy for in-plane and out-of-plane polarizations within RPA+PBE approach. The inset shows a comparison of optical conductivity for the in-plane polarization in the visible range as a function of wavelength including graphene (black line)}
\label{fig07}
\end{center}
\end{figure}

We next discuss the optical conductivity of these 2D systems. 
The real part of the optical conductivity is related to $\mathrm{Im } [\epsilon_{\alpha \beta} (\omega)]$ by \cite{wooten1972optical}:
\begin{equation}
\mathrm{Re } [\sigma_{\alpha \beta}(\omega)] = \frac{\omega}{4 \pi} \mathrm{Im} [\epsilon_{\alpha \beta} (\omega)].
\end{equation}
The real part of the optical conductivities are presented in Fig. \ref{fig07}, where we also include an inset in order to compare with pristine graphene in the visible range of light  \cite{Shahrokhi2016,Shahrokhi2016a}. 
The optical conductivities begin with a gap, which is due to the semiconducting properties of these nanosheets. 
The first prominent optical conductivity peaks occur at 1.95, 1.57, 1.42 and 1.69 eV for BC$_3$, BC$_6$N-1, BC$_6$N-2 and C$_3$N, respectively. 
Meanwhile, the main peaks of optical conductivities are located at at 11.84, 12.31, 12.78 and 13.16 eV, respectively. 
The main peak in $\mathrm{Re } [\sigma_{\alpha \beta}(\omega)]$ for out-of-plane polarization locate at energy levels between 9.55 and 13.59 eV. 
Furthermore, it can be seen that the optical conductivities of all studied nanosheets in the visible range of light are higher than that of the graphene. 
Notably, BC$_3$ and C$_3$N monolayers present very strong conductivity in the 500--700 nm range along with very high optical conductivities. These enhancements in the optical conductivities further highlight the desirable performances of BC$_3$ and C$_3$N nanosheets for applications in photovoltaic cells.

\begin{figure}[htbp]
\begin{center}
\includegraphics[width=0.9\linewidth]{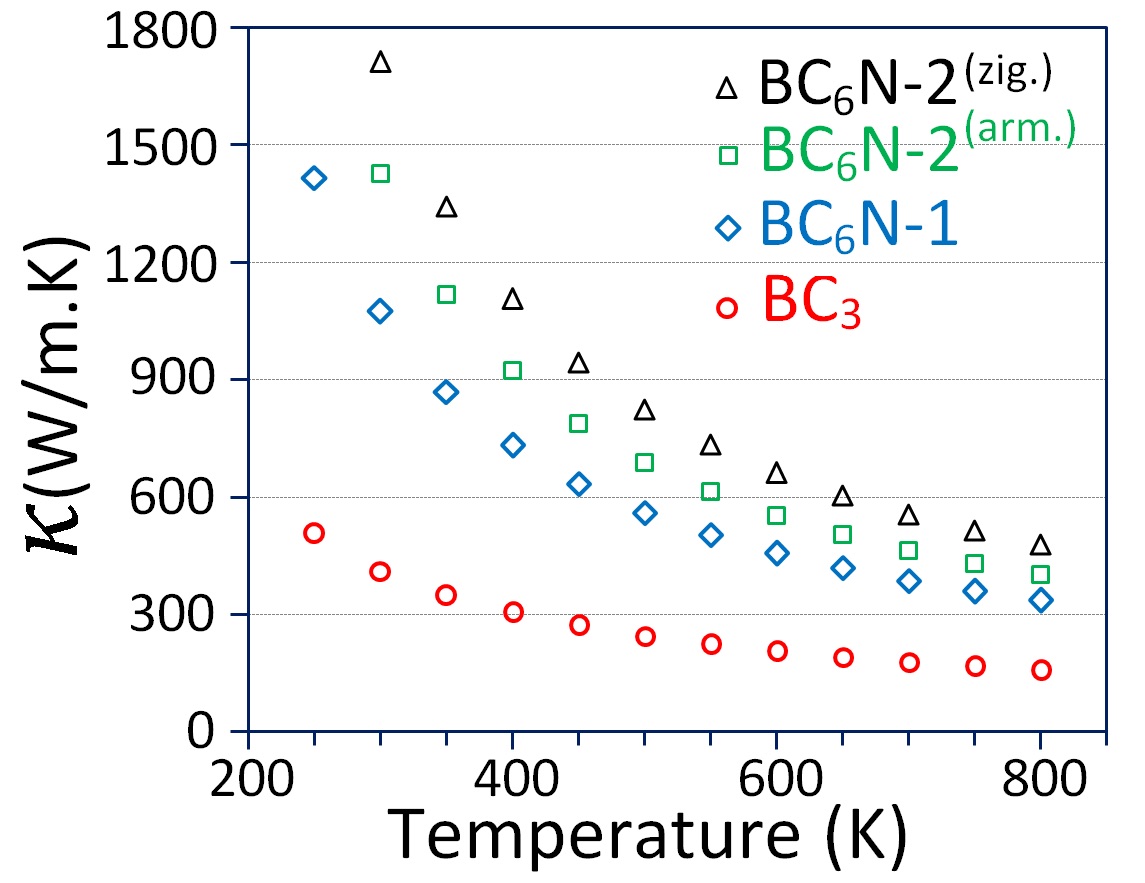}
\caption{Lattice thermal conductivity of BC$_3$, BC$_6$N-1 and BC$_6$N-2 (armchair and zigzag directions) calculated from first principles lattice dynamics. The room temperature conductivities are 410 W/m.K for BC$_3$, 1080 W/m.K in the case of BC$_6$N-1, 1430 and 1710 W/m.K along the armchair and zigzag directions of BC$_6$N-2, respectively.}
\label{fig08}
\end{center}
\end{figure}

Last but not least, we consider the thermal conductivity of BC$_3$, BC$_6$N-1 and BC$_6$N-2 monolayers calculated from first principles. 
Since these materials are semiconductors, phonons are their major heat carriers, such that we focus our predictions on the lattice thermal conductivity of these graphene-like monolayers. 
According to our calculations, the diagonal elements of the thermal conductivity tensor relative to the in-plane directions are identical within the precision of the method, with the notable exception of BC$_6$N-2, which presents anisotropic thermal conductivities.
In all cases, the conductivity tensor element corresponding to the out-of-plane direction vanishes. 
Therefore, we can assert that thermal transport in BC$_3$, BC$_6$N-1 is isotropic along armchair and zigzag  directions, in agreement with C$_3$N, C$_2$N \cite{Mortazavi2016a} and graphene \cite{Pereira2013}, while BC$_6$N-2 presents anisotropic thermal transport characteristics.
In Fig. \ref{fig08} we show the lattice thermal conductivities as a function of temperature in the range 250 to 800 K. 
In the case of BC$_3$ and BC$_6$N-1 we present the average of the non-vanishing diagonal terms of the conductivity tensor, while for BC$_6$N-2 we show both in-plane components separately, which are related to the armchair and zigzag directions.
In the temperature interval considered, the conductivity of BC$_6$N-2 in either direction is always larger than that of BC$_3$ and BC$_6$N-1. 
The room temperature lattice thermal conductivities are 410 W/m.K for BC$_3$, 1080 W/m.K in the case of BC$_6$N-1 (the structure without B-N bonds), 1430 and 1710 W/m.K along the armchair and zigzag directions of BC$_6$N-2 (the structure with B-N bonds), which are remarkably high for monolayers. 
For reference, the room temperature thermal conductivity of bulk copper is 400 W/m.K, while for hexagonal BN it is 600 W/m.K \cite{Lindsay2011a}, and for graphene we have 2900 $\pm$ 100 W/m.K \cite{Fan2017}.
It is worth of notice that we have previously observed a strong correspondence between anisotropies in elastic moduli and thermal conductivities \cite{Pereira2016,Mortazavi2016a,Mortazavi2017a}.
In several of our previous studies, 2D materials with isotropic elastic modulus also presented isotropic thermal conductivities.
BC$_6$N-2 monolayers seem to be a notable exception, since it presents isotropic elastic moduli but different values for thermal conductivity along in-plane directions.

\begin{figure}[htbp]
\begin{center}
\includegraphics[width=0.9\linewidth]{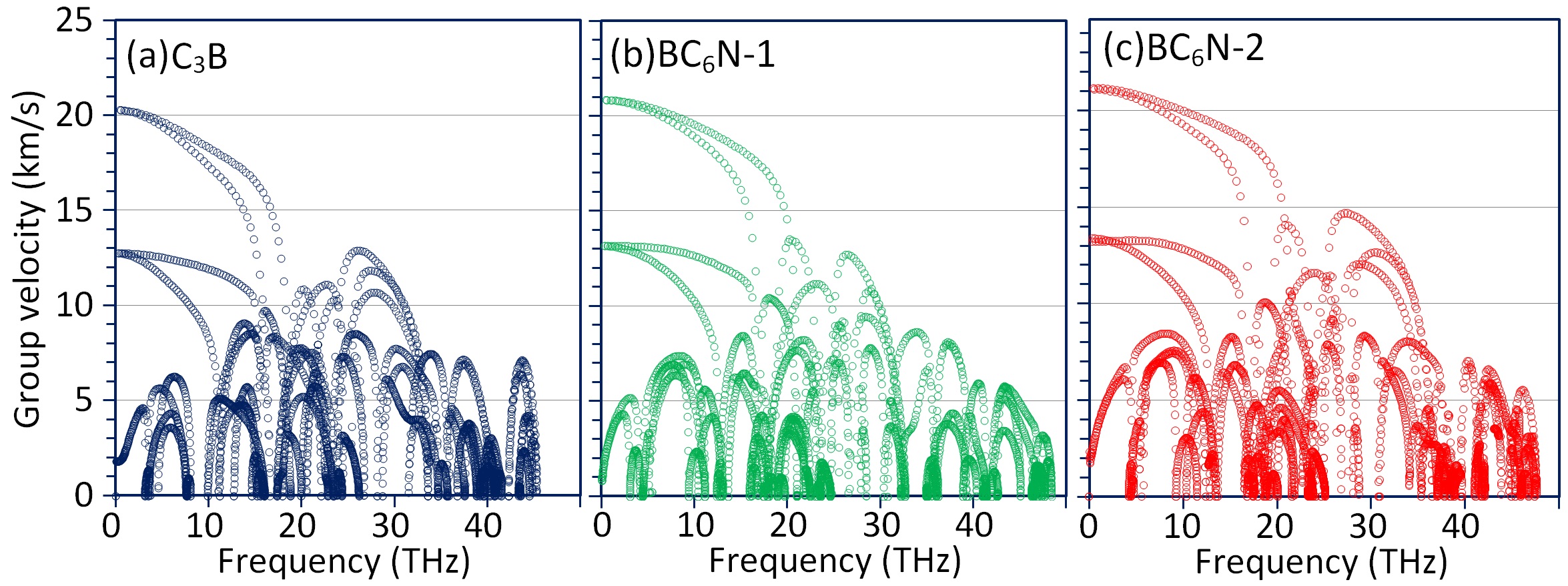}
\caption{Phonon group velocities of BC$_3$, BC$_6$N-1 and BC$_6$N-2 calculated from first principles lattice dynamics. On average he group velocities are higher in the case of BC$_6$N-2, consistent with its higher thermal conductivity.}
\label{fig09}
\end{center}
\end{figure}

At first it might seem counterintuitive for BC$_6$N-2 to present higher lattice thermal conductivities relative to BC$_3$ and BC$_6$N-1, and in order to clarify this matter we look at the phonon group velocities shown in Fig. \ref{fig09}. 
The major contribution to the thermal conductivity of both materials comes from their acoustic phonon modes. 
If we focus on the lower frequency range of the acoustic phonons, say below 15 THz, it is noticeable that on average  the group velocities are higher in the case of BC$_6$N-2, consistent with its higher thermal conductivity.
This comparison of phonon group velocities has been enough to understand why a certain material presents a larger thermal conductivity relative to another material in some of our previous works \cite{Mortazavi2016, Mortazavi2017b}. Nonetheless, it is not always possible to atribute the difference to group velocities alone, and other quantities migth need to be considered such as phonon mean free paths or scattering rates \cite{Peng2018}.

In general, doping of a pristine material decreases the thermal conductivity due to an increase in phonon scattering rates.
However, our results suggest that in the case of BC$_3$ and C$_3$N nanosheets, controlled chemical doping and replacement of a single B or N atom in every unit-cell can substantially improve the thermal conductivity.
Furthermore, for these nanosheets, controlled doping also modulates the electronic structure by creating direct bandgap semiconductors. 
Therefore, our results reveal an unusual but very promising finding. 
We predict the possibility of opening a direct bandgap in graphene with minimal sacrifice of its ultrahigh thermal conductivity via co-doping with B and N atoms.

\section{Concluding remarks}

Graphene-like BC$_3$ and C$_3$N are among the most attractive carbon-based 2D semiconductors. Motivated by the outstanding properties of C$_3$N and BC$_3$ nanosheets, we propose two novel graphene-like semiconductors with BC$_6$N stoichiometry. 
We conducted extensive density functional theory calculations to explore the mechanical properties, electronic structure, optical characteristics and thermal conductivity of free-standing single-layers of BC$_3$ and BC$_6$N. 
Phonon dispersions of BC$_3$ and BC$_6$N nanosheets were found to be free of imaginary frequencies, and AIMD simulations at 1000 K confirm their mechanical stability. 
First-principles calculations reveal that BC$_3$ and BC$_6$N monolayers present isotropic and outstandingly high elastic moduli of 256 and 305 N/m, respectively.
 BC$_3$ and BC$_6$N nanosheets exhibit higher tensile strength and stretchability along the zigzag direction when compared to the armchair direction. The maximum tensile strength of BC$_3$ and BC$_6$N were predicted to be 29 and 33.4 N/m, respectively, only around 30\% lower than that of the graphene. 

Our analysis of electronic and optical characteristics of BC$_3$ and BC$_6$N monolayers also reveal promising physical properties.
Notably, BC$_3$ and BC$_6$N monolayers, without and with B-N bonds, show indirect and direct bandgaps, respectively, with values of 1.82, 2.10 and 1.77  eV, according to the HSE06 functional. 
Finally, the optical response of these graphene-like materials, including the imaginary and real part of dielectric function, absorption coefficient and optical conductivity for in-plane and out-of-plane polarizations were investigated. The first absorption peaks along the in-plane polarization reveal that these novel 2D nanostructures can absorb visible light, suggesting their prospect for applications in optoelectronics and nanoelectronics. 
Moreover, the absorption coefficient and optical conductivity of these nanosheets in the visible range were observed to be larger than those of graphene. 

As an exciting finding, the room temperature lattice thermal conductivity of  BC$_3$ and BC$_6$N monolayers were predicted to be remarkably high at 410  and 1710 W/m.K, respectively, highly desirable to contend with heating dissipation concerns.
Our extensive first-principles calculations highlight the outstanding physical properties of graphene-like BC$_3$ and BC$_6$N nanosheets, and suggest them as strong and highly thermal conductive semiconductors, suitable for the design of advanced electronic, optical, energy storage and thermal management devices.

\section*{Acknowledgments}
B.M. and T.R. acknowledge financial support from the European Research Council for the COMBAT project (Grant no. 615132).
B. M. and X. Z. particularly appreciate funding by the Deutsche Forschungsgemeinschaft (DFG, German Research Foundation) under GermanyÕs Excellence Strategy within the Cluster of Excellence PhoenixD (EXC 2122, Project ID 390833453).
L.F.C.P. acknowledges financial support from Conselho Nacional de Desenvolvimento Cient\'ifico e Tecnol\'ogico (CNPq) for the project ``Thermal and electronic transport in 2D materials" (Grant no. 309961/2017).

\section*{References}

\bibliography{/Users/pereira/Documents/library}

\end{document}